# Applications of Multi-view Learning Approaches for Software Comprehension

## Amir M. Saeidi[a], Jurriaan Hage[a], Ravi Khadka[a], and Slinger Jansen[a]

a    Department of Information and Computing Sciences, Utrecht University, The Netherlands

**Abstract**    Program comprehension concerns the ability of an individual to make an understanding of an existing software system to extend or transform it. Software systems comprise of data that are noisy and missing, which makes program understanding even more difficult. A software system consists of various views including the module dependency graph, execution logs, evolutionary information and the vocabulary used in the source code, that collectively defines the software system. Each of these views contain unique and complementary information; together which can more accurately describe the data. In this paper, we investigate various techniques for combining different sources of information to improve the performance of a program comprehension task. We employ state-of-the-art techniques from learning to 1) find a suitable similarity function for each view, and 2) compare different multi-view learning techniques to decompose a software system into high-level units and give component-level recommendations for refactoring of the system, as well as cross-view source code search. The experiments conducted on 10 relatively large Java software systems show that by fusing knowledge from different views, we can guarantee a lower bound on the quality of the modularization and even improve upon it. We proceed by integrating different sources of information to give a set of high-level recommendations as to how to refactor the software system. Furthermore, we demonstrate how learning a joint subspace allows for performing cross-modal retrieval across views, yielding results that are more aligned with what the user intends by the query. The multi-view approaches outlined in this paper can be employed for addressing problems in software engineering that can be encoded in terms of a learning problem, such as software bug prediction and feature location.

**ACM CCS 2012**

- **Software and its engineering** → **Software reverse engineering**; *Maintaining software*;
- **General and reference** → *Evaluation*;
- **Information systems** → *Clustering*; *Collaborative filtering*; *Nearest-neighbor search*; *Similarity measures*;
- **Computing methodologies** → *Information extraction*;

## The Art, Science, and Engineering of Programming

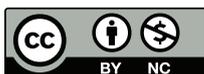

 **Introduction**

Program comprehension is the process of understanding a program's meaning and behavior to perform software evolution. Understanding a program requires attaining a mental map of the program. This highlights the importance of developing program comprehension techniques, built in development environments as integrated tools to facilitate different phases of software development. These techniques range from extracting a high-level overview of the structure of the software to tools for performing code search and refactoring. Classic program comprehension techniques are dominated by the formal, or logico-deductive approach [4], where formal analysis tools such as data flow analysis and type inference are used to deliver meaningful representations for understanding the program. Such software analyses treat a program as a mathematical object with formal semantics, essentially working on the abstract syntax tree to perform the analysis.

Developers embed information about the specification of a program in other forms as well, including the documentation, identifier names and comments, commit history, test cases, etc. Michael D. Ernst in [22] argues that *"Traditionally, programming language researchers have viewed it [software systems] as an engineered artefact with well-understood semantics that is amenable to formal analysis. An alternative view is as a natural object with unknown properties that has to be probed and measured in order to understand it."*. Machine learning provides the means to develop software analysis techniques that exploit this naturalness of software systems to mine statistical patterns that characterizes the software. These techniques can augment existing software analyses or replace them all together, ranging from software modularization, summarizing code as text, i.e. code summarization [25] to ranking the results of code search [5], feature location [18] and ranking the code completion suggestions [3, 41].

Software analysis tools need to integrate information from various artefacts that developers create to infer useful information. These artefacts include test cases, source code identifier names, call graphs, module dependencies, program structure, evolution history in a version control repository and the execution logs. Muti-view learning allows for systematic integration of information from different modalities about an object to allow for inference of useful information from them. Our notion of view is defined as a domain-based perspective of a software system (also known as a variable group or representation scheme), in which it 'may' be easy to capture the notion of similarity in each view. So, the concept of multi-view learning is concerned with combining different sources of information (also known as *data fusion* or *data integration*) to reduce the noise in each view, as well as leverage the interactions and correlation between information spaces to obtain a better and more refined understanding of the structure of the system. We will employ different multi-view learning techniques to perform the learning tasks: 1) multiple kernel learning, 2) co-training, and 3) subspace learning. In this paper, we demonstrate the benefits of multi-view learning for three applications, 1) software modularization, 2) component-level recommendations, and 3) cross-modal software search. We make a quantitative comparison of the aforementioned techniques for the first two tasks, while we conduct a qualitative evaluation of code search across views using subspace learning.

Similarity computation is a crucial step of any learning algorithm. Coupling heuristics[1] are proposed to capture the similarity between software entities at different abstraction levels. Most common coupling heuristcs are based on data and control dependencies. In addition to such relationships, there are also coupling metrics defined based on conceptual relationship between software components. Although each coupling heuristic tries to capture some design decision at some abstraction level, there is no consensus [2, 11] as to which heuristic is the dominant factor in capturing the similarity between software entities. Each coupling heuristic defines its own notion of similarity/dissimilarity in a particular view, such as the evolutionary information or the call dependency graph. Each view may contain information that is unique; therefore, multiple views can be employed to more comprehensively and accurately describe the data.

The selection of variable group and the representation schemes play a central role in learning similarity in each view of software system. Classical approaches depend on feature extraction for computing the similarity. For a successful feature extraction from different knowledge bases associated with the software system (such as the revision history or the source code corpus), not only is extensive domain knowledge required but a considerable amount of knowledge is lost during this process. Kernel-based methods to learning are an alternative to explicit feature extraction. The main idea is to implicitly map a problem into another space that is more suitable for finding the solution. We will explore different kernel-based methods for learning on different artefacts of the software system and devise a suitable notion of similarity for that domain.

Our contribution in this paper is as follows:

- We establish good choices for kernel-based learning for source code analysis in each view.
- We present state-of-the-art techniques from multi-view learning to perform program comprehension tasks.
- A comparative study of the single-view and multi-view techniques is performed on a set of 10 relatively large Java open source projects for both clustering as well as recommendation.
- We perform a qualitative evaluation of cross-modal retrieval on the jEdit[2] project using cross-view subspace learning.

We first survey related work on clustering of software systems and recommender systems as well as source code search techniques in Section 2. In Section 3, we describe different approaches to multi-view learning. We proceed in Section 4 by encoding various coupling heuristics in terms of a learning problem for capturing similarity between software entities, and explore different kernel algorithms for doing so. In Section 5, we proceed by making a comparison between different approaches to multi-view clustering and cross-domain recommender systems, and establish good choices of kernels for learning from different software artefacts for each task. A qualitative evaluation of source code search is given using cross-modal retrieval. We conclude in Section 6 and outline future work.

---

[1] Also known as coupling metrics or measurements.
[2] http://www.jedit.org/ (visited on 2019-01-29).

## 2 Related Work

We divide the related work into three categories: 1) single-view and hybrid techniques for software modularization, 2) the techniques for recommendation in software engineering, and 3) various approaches to source code search.

### 2.1 Software Modularization

The majority of the software clustering approaches presented in the literature attempt to modularize software systems by analysing structural dependencies between source code entities. The high-cohesion, low-coupling modularity principle is encoded in terms of a search problem in [33], where an objective function called Modularity Quality (MQ) is optimized. MQ measures both the total dissimilarity between the source code units in different clusters as well as the total similarity within the same cluster. Praditwong and Yao [40] formulate the high-cohesion, low-coupling metric in terms of a multi-objective optimization problem. They employ a genetic algorithm to compute the Pareto optimal cuts of the graphs and demonstrate that the produced results are significantly better for both weighted and unweighted graphs, compared to single-objective encoding of the MQ problem.

Semantic similarity (also known as lexical similarity) is a coupling heuristic employed in [29] to perform *semantic clustering*. The authors borrow techniques from the information retrieval field to partition the system based on a common use of vocabulary. An observation they make through performing semantic clustering on two Java open source projects is that semantic similarity fails to fully capture the inherent structural architecture of the system. In this approach, the same weight is given to the vocabulary extracted from different parts of the source code. In [16, 17], a weighing mechanism is introduced to assign different importance to the information extracted from different zones in the source code such as comments or identifier names. The authors show that this approach further improves the quality of semantic clustering. Conceptual metrics and semantic clustering are employed in [45] to investigate whether latent topics in the source code explain the intention behind a software re-organization. The study conducted on six real-world re-modularization projects demonstrate that the latent topics in the source code improve through restructuring of the systems.

Recently, more attention has been paid to integrating different sources of information relevant to software system to improve software modularization. Andritsos and Tzerpos [6] combine non-structural attributes such as timestamps and ownership with structural dependencies and show that this approach results in improvement in the quality of software clustering. Beck and Diehl [10] make a comparison between structural dependencies and evolutionary coupling and show that by integrating these coupling concepts, the quality of clustering results can be improved. In [9], latent topics and structural dependencies of the classes in a package are integrated to identify semantic relationships between them and those relationships are used to restructure the packages. The authors show that by combining semantic and structural couplings, it is possible to improve software modularization. This problem is encoded in terms of a search problem in [42] to search for a partition that maximizes hypotheses in each respective view. The hypotheses in each view are competing for the outcome of the optimization problem.

## 2.2 Recommendation systems in software engineering

RASCAL [35] is a Collaborative Filtering based recommender system that tries to predict the next method a developer may use, based on the methods used in similar classes. The similarity between any two classes is computed based on the methods they call. The predictions given for the call graph in our work is similar to this technique, however, we focus on component-level dependencies between different classes.

An approach for automating 'Extract Class' refactoring is proposed in [8] that is implemented as a plug-in for eclipse. The authors represent a class in terms of a weighted graph, where each node corresponds to a method, and the weights on edges correspond to structural and semantic relationship. They employ techniques from graph theory to use the closeness measure between methods to extract new classes with higher cohesion. The authors build on their previous work to propose an approach for performing 'Move Method' refactoring, based on semantic and topical information [7]. Although our approach does not give fine granular source code refactorings, it exploits the diversity of information spaces of a software system to give high-level recommendations through structured integration of data.

A search-based multi-objective formulation of refactoring recommendations is introduced in [39]. The multi-criteria based on which the refactorings are suggested are (i) improving design quality, (ii) preservance of the design coherence after performing refactoring, (iii) minimizing code changes, and (iv) maximizing the consistency with change history. They evaluate their approach on 6 open source projects and conduct an industrial validation of their method.

## 2.3 Source Code Search

Code search takes a query specified in some form to find relevant pieces of code in the software system. Code search methods fall into three categories: Natural, Syntactic and Semantic. The natural code search uses Natural Language Processing (NLP) techniques to retrieve code segments that match the specification. Syntactic code search uses syntactic features such as keywords and identifier names (e.g. method and variable names) as the specification. On the other hand, semantic search engines leverage semantic information in the software system to improve search results. Exemplar is a search engine that uses a keyword search that returns applications based on API calls and the flow of data among those APIs [24]. Portfolio [36] is a semantic code search system which locates chains of functions that implement a given query and visualizes the dependencies of those functions. An input-output example-based semantic code search is proposed in [49] which uses an SMT solver to identify programs or program fragments which behave as the programmer-provided specification. The aim of cross-modal retrieval is to integrate information across various artifacts of software system including the keywords, type hierarchy and API usage to both find software entities across modes of data, but also improve the quality of ranked results in each view.

## 3   Multi-view Learning for Software Analysis

There are many problems in which different views of data are available. For example, a web document can be represented by its url links and the words appearing on the page, while multilingual documents have a representation in each language. A software system can also be viewed from different perspectives, corresponding to a variable group or representation scheme describing that domain. For example, evolutionary information is a set of group variables describing the co-changing frequency of software units.

Traditional learning methods such as clustering and classification, operate on a single view, but recently more attention is being paid to learning from multiple views simultaneously. In contrast to single view learning, multi-view learning jointly optimizes all the functions to exploit the complementary information in each view to improve the learning performance [51]. We employ various multi-view learning techniques to combine different sources of information to help improve the quality of software comprehension tasks, such as modularization, recommendation and source code search.

We make two fundamental assumptions that are key to the success of multi-view learning:

1. Sufficiency: each view is sufficient for learning on its own. For instance, an image of a plane or an eagle that have the same colour histogram are not sufficient to distinguish between the images.
2. Compatibility: the target function of both views predict the same labels for co-occurring features with a high probability.

Although it is possible to perform multi-view learning on datasets with incomplete bipartite mappings between different views [21], in this paper, we will assume the source code units have a complete representation in each view.

The general approach to performing a learning task in multi-view setting is depicted in figure 1. The data collected from each information space is used to compute similarity between software entities. These similarities highly depend on the domain, and a suitable similarity function needs to be selected that best captures similarity in that domain, as will be explained in Section 4. Once similarity matrix for each view is constructed (shown as *View Similarity* in figure 1), they are fed into the *data fuser* to combine the information that is unique in each view, together which can help better understand the underlying dynamics in each view. The jointly learned space(s) can then be used to perform a particular learning activity, including regression, classification or clustering. In the following, we outline the multi-view learning algorithms used in this paper. We proceed by briefly describing some of the program comprehension

■ **Figure 1** General approach to data fusion for conducting multi-view learning

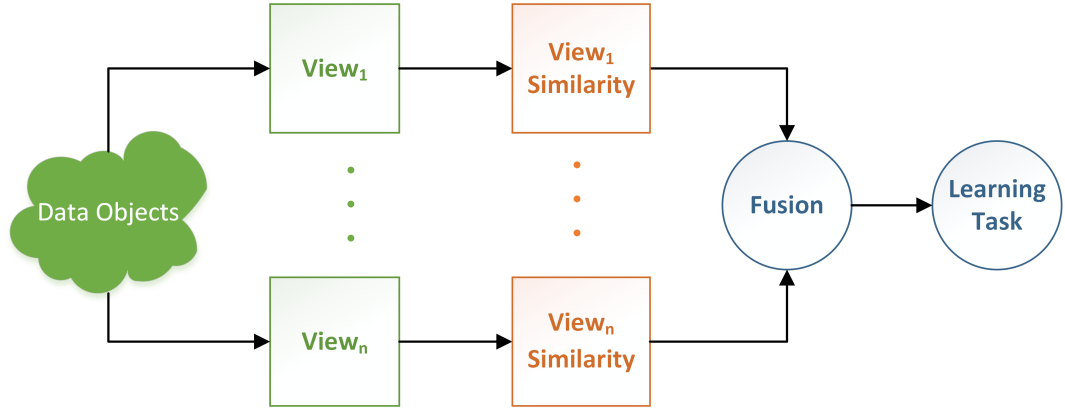

■ **Table 1** A brief comparison of multi-view learning approaches

| Category | Representatives | Main Applications | Description |
|---|---|---|---|
| MKL | Kernel addition, Kernel product | clustering, semi-supervised learning | Kernels of each view are combined using a kernel-based method |
| Co-training | co-training [14], co-EM [38], co-clustering [52] | semi-supervised learning, transfer learning | Separate learners on distinct views are trained by forcing them to be consistent across views |
| Subspace learning | CCA [26], KCCA | dimension reduction | Learn shared subspace across views by assuming they are generated from a latent view |

tasks that can be encoded as a learning problem, that will be investigated throughout this paper to make a quantitative comparison between single-view and multi-view approaches.

### 3.1 Multi-view learning Approaches

Multi-view learning algorithms can be classified into three categories: 1) Multiple kernel learning (MKL), 2) Co-training , and 3) Shared subspace learning. Table 1 gives a brief comparison of the three classes of multi-view learning. MKL combines the kernel representation of each view by means of a linear or non-linear combination to yield a unified kernel representation. Co-training based algorithms such as co-EM, and co-clustering try to jointly maximize the mutual agreement by training alternatively between two distinct views of the data. Subspace learning techniques such as CCA and KCCA learn a latent subspace, shared by different views. The main assumption is that the observed views are generated from the same latent subspace. The shared subspace can then be used to perform a learning task. For more information about various multi-view learning approaches, please refer to appendix A.

### 3.2 Modularization

Modularization, also known as software clustering, refers to the re-location of software units in a cluster which is more cohesive and aligns better with a specific functionality of the system. Software clustering is used to enhance the comprehension and maintainability of a software system. Multi-view clustering leverages abundance of information in each view to recover the underlying architecture. The main assumption that is essential to success of multi-view clustering is that the interactions and correlations within each view is similar. We formulate the software modularization problem as a multi-view clustering to study whether any benefits can be gained by data integration for this particular task.

### 3.3 Collaborative Recommendations

Recommender systems are a prevalent technique in search engines, social networks and products in general, that aim to present most desirable information to users based on the item 'ratings' or user 'preferences'. Collaborative Filtering (CF) is a well-known technique in recommender systems, which tries to exploit the relationship between users, and recommends items based on the preference of similar users. In this work, we focus on collaborative-recommendations for software system units. The basic idea is to find the source code units similar to a particular module, and use the information items present in those units to recommend item information to the module. The type of recommendations depends on the target information space. For change history, the problem amounts to predicting whether a source code unit should also be committed with other modules. For a graph structure like the call graph, this problem amounts to predicting the likelihood of the existence of a link (or reliability of a link), commonly known as the link prediction problem.

   Due to heterogeneity of data sources, multi-view recommender systems (sometimes, called cross-domain recommender systems) take advantage of unique and complementary information among related domains to recommend items in a target domain [15]. Multi-view recommender systems can improve the accuracy of recommendations by transferring knowledge across the domains. Furthermore, multi-view recommendations can overcome one of the most important problems in recommender systems, namely, the *Cold Start* problem. When a new user starts using the system, or a new product item is added to the catalogue, as no history information exists about the object, it is very difficult to give reliable recommendation about the object. Multi-view recommender systems try to alleviate cold-start problem by using the information from one domain to give recommendation in another domain, where no priori knowledge about the object exists. We aim to investigate multi-view collaborative filtering recommender systems in the context of program comprehension, by comparing their performance with single-view recommendations to understand how well they work, as well as identify the domains in which these systems perform the best.

### 3.4 Cross-modal Retrieval

The goal of cross-modal retrieval is to take one type of data (e.g. image or text) as the query to retrieve relevant data of another type. Furthermore, when users submit a query of a particular type of data, they can obtain more robust search results, given that different modalities of data can provide complementary information to each other. By combining different modalities of the data, we can reduce the semantic gap, thus, obtaining higher similarity for objects that are semantically similar.

Existing search engines for software systems heavily rely on the textual content of source code units to perform a search, whether those are natural text appearing in comments or syntactic features like identifier names. The similarity between a query and a source code unit is defined based on their textual feature vectors. Textual relevance models including Vector Space Model and Language Models may not precisely capture what the user intends by the query. For instance, making a query for 'UI Manager' will find modules that have exact or near matches of the query in their textual representations, returning modules that have common terms such as 'manager' and 'UI'. In absence of those terms, no match will be found. However, the intent of the user by 'UI Manager' is those set of classes which act as managing the lifecycle of some set of UI elements, such as pop-up and modal dialogs.

A good search engine should be able to abstract away from textual descriptors of a query to find what the user actually means by the search query. Multi-view subspace learning enables semantic abstraction, i.e., the representation of source code from textual perspective to how they interact in other representations in terms of low-level semantic descriptors. This makes the approach well suited for nearest neighbor methods, which find the most similar objects in the low-level representation (i.e. embedding). We will employ cross-modal retrieival to explore the quality of search results obtainable through subspsace learning in context of jEdit project.

## 4 Kernels for Encoding Coupling Heuristics

For software systems, various coupling heuristics are proposed to capture similarity between software units. Each coupling heuristic in each view tries to explain what it means for two source code units to be similar or dissimilar. The notion of similarity heavily depends on the domain a coupling heuristic is defined for. Kernel-based methods are one such technique that induce similarity measures in different input spaces (views). Our primary objective is to investigate the extent to which different kernel functions capture similarities between source code entities. We first introduce kernel-based methods for learning and then focus our attention on three coupling heuristics and encode them in terms of kernel-based learning problems.

### 4.1 Kernels

A function $k(.,.) : \Omega \times \Omega \rightarrow \mathbb{R}$ for some input space $\Omega$ and real numbers $\mathbb{R}$, denotes a notion of similarity such as distance or proximity between two objects $x$ and $y$ in $\Omega$. A kernel function or positive definite function [46, 47] enjoys two important properties that are essential when used to compute similarity in different types of input spaces: 1) computing the kernel between any two objects amounts to computing the inner product of some transformation of the objects $x$ and $y$ in some other high-dimensional space, without computing the exact mapping from the input space into this new feature space (this property is also known as the *kernel trick*), 2) kernel functions can be used to compute the similarity between structured objects such as graphs that cannot be expressed in terms of a set of features.

For each kernel function $k(.,.)$, a *kernel matrix* is defined as a $n \times n$ symmetric kernel matrix denoting the computation of the pairwise kernel function between objects $x_i$ and $x_j$. It satisfies the property that it is a symmetric positive semi-definite matrix[3].

There are various types of kernels described in [47] which operate on general types of data, some of which operate on a vector space while others work on graph structures. A useful kernel is expected to capture an appropriate measure of similarity between two objects $x$ and $y$, specialized for a particular domain. The choice of kernel is driven by the geometry of the problem, and some of the kernel functions have a hyper-parameter that needs to be tuned. In this paper, we focus on the kernels that are suitable for the domain of source code analysis with a focus on different artefacts in the software systems. We will explore state-of-the-art kernels for vector spaces, strings as well as graph structures to identify a suitable kernel for learning on software artefacts, and study the effect of varying values for the parameters. For more information about the kernels used in this paper, please refer to appendix B.

## 4.2 Coupling Heuristics

### 4.2.1 Structural Modularity

Structural dependencies denote a notion of dependency between source code units such as a call dependency or an inheritance relationship. Two classes are coupled if there is a structural dependency between them. Based on this heuristic, two software units are similar if they are structurally coupled together. We have limited ourselves to the call dependency graph, however our technique can be used to incorporate other structural dependencies.

### 4.2.2 Evolutionary Coupling

Version Control Systems (VCS) store information about the evolution of a software system that can be used to better capture the intrinsic structure of the system. Useful information stored in a repository include code change patterns (e.g., how often two software artefacts are changed together), implementation decisions associated with specific changes as potentially described in commit messages, and other auxiliary information about the authors of the change. Beck and Diehl [11] call this coupling, *evolutionary coupling*. We have followed the approach by Zimmermann and Weißger-ber [56] to extract the evolutionary information from the transaction history. We collect the entire transaction history for the tagged release version used in the dataset, by means of standard tooling that comes with SVN or git, depending on the type of version control system used by the project. For each transaction, we identify the files that have been added, deleted, or modified as the set of files that have co-changed during one transaction. We have omitted transactions with larger than 30 files to reduce noise. We will employ the polynomial and the Gaussian kernels to construct a similarity matrix for the evolutionary coupling.

---

[3] Not all similarity matrices are positive semi-definite but in this paper, we will only deal with kernel matrices.

### 4.2.3 Lexical Similarity

Lexical similarity, or semantic similarity involves interpreting the source code of software system as mere plain text, where common use of vocabulary indicates a coupling between the modules. Computing the lexical similarity based on the text of the source code requires some kind of similarity measure between strings. The kernel functions used here for computing the similarity, as in text classification are: 1) the bag-of-words kernel, 2) the polynomial kernel, 3) the RBF (Gaussian) kernel, and 4) the string kernel.

For vector space kernels, we have followed the information retrieval method, Latent Semantic Indexing (LSI) [20] as outlined in [29] for software analysis to build the feature representation of the source code corpus (i.e. bag-of-words). Constructing the bag-of-words involves applying a preprocessor to extract terms from the source code corpus, followed by breaking-up composite terms using standard naming convention in Java i.e. CamelCase. It proceeds by eliminating common terms that occur in a natural language, such as "the" and "is", as well as reserved keywords in a programming language, including "public" and "class". A stemming algorithm is then used to extract the root of words that are derived from a common radix. For instance, words "stems", "stemmer" and "stemming" are reduced to the root word "stem". We proceed by applying term frequency–inverse document frequency *tf-idf* weighing mechanism, which punishes words that are prevalent throughout the corpus, such as "get" and "index", while increasing the weight of less-frequent words. We first project the bag-of-words into a lower dimensional space before applying the kernel functions. We have adopted the formula suggested in [29] to compute the number of dimensions formulated as $r = (m * n)^{0.2}$, where $m$ and $n$ are the number of words and documents, respectively. LSI helps with reducing noise associated with synonymy and to a lesser extent polysemy in the text of the source code.

Before applying the string kernel, as we do in the bag-of-words normalization process, we will take the text of the source code corpus through a preprocessing phase involving stemming, and stop-word removal (natural language common words as well as programming language reserved words).

Table 2 lists the kernel algorithms used for different input spaces. Most kernel functions need a value for some parameters (given in the "Par" column). The values tested (see the "Tested values" column) were tuned in the experiments, based on the accuracy they have achieved for a particular learning problem. Please note that the best model fits cannot be generalized, and a particular model performing well for a learning task does not necessarily perform well for another task. Kernel-based

■ **Table 2** The kernels for different input spaces with the values tested for their parameters

| Name | Abbr. | Eqn. | Par. | Tested values |
|------|-------|------|------|---------------|
| Bag of words | $\mathbf{K}_{BoW}$ | 7 | - | - |
| Polynomial | $\mathbf{K}_{Poly}$ | 5 | $d$ | $1, 2, 3, 4, 5$ |
| Gaussian | $\mathbf{K}_{RBF}$ | 6 | $\alpha$ | $10^{-5}, \ldots, 10^2$ |
| Constant | $\mathbf{K}_{Cons}$ | 8 | - | - |
| p-spectrum | $\mathbf{K}_{Spec}$ | 8 | $p$ | $1, 2, 5, 10, 15, 20$ |
| Exponential decay | $\mathbf{K}_{Exp}$ | 8 | $\lambda$ | $1, 2, 5, 10, 15, 20$ |
| Exponential diffusion kernel | $\mathbf{K}_{ED}$ | 9 | $\alpha$ | $10^{-5}, \ldots, 10^2$ |
| Laplacian exponential diffusion kernel | $\mathbf{K}_{LED}$ | 10 | $\alpha$ | $10^{-5}, \ldots, 10^2$ |

algorithms are not suitable when dealing with large-scale dataset (millions of objects) as they involve a matrix inverse or factorization. However, when dealing with software systems, they are a good choice for computing the similarity matrix. In general, software systems consist of no more than a 1000 software units.

## 5 Empirical Evaluation

In this section, we report results on the accuracy of kernels computed on different artefacts of a software system for multi-view clustering and recommendation. We then proceed by using the established choices of kernels to perform a qualititative evaluation of cross-modal source code search. We aim to investigate the benefits of data fusion in the performance of the learning problem as well as identify the views that align best with those implicitly embedded by the developers[4]. We first present the benchmark for evaluating different kernel-based learning methods.

### 5.1 The Dataset

The dataset presented in Table 3 comprises 10 open source Java projects. We have used the tool *Chord* [37] to construct the call dependency graph of these systems. Constructing the call graph requires specifying the entry points to the system. Any class deemed unreachable through the entry point is removed. Hence, inclusion of the classes is limited by our knowledge about different configurations for running the system. Furthermore, in Table 3 information about the number of call dependencies and number of transactions for each project is given. The information corresponds to the datasets after intersection of all views on software units.

---

[4] All the implementations of the multi-view clustering and collaborative recommendation can be found as part of the GeLaToLab at https://github.com/amirms/GeLaToLab (visited on 2019-01-29).

■ **Table 3** The benchmark for empirical evaluation of multi-view learning approaches

| System | Description | Version | #Classes | KLOC | #Calls | #Trans. |
|--------|-------------|---------|----------|------|--------|---------|
| Ant | building library | 1.9.3 | 96/691 | 31.07 | 325 | 952 |
| Hadoop | distributed computing library | 0.20.2 | 361/707 | 102.53 | 395 | 503 |
| JDOM | XML library | 2.0.5 | 63/140 | 21.93 | 154 | 130 |
| JDT Core | Java Development Tools | 3.8 | 356/1276 | 162.59 | 2661 | 3043 |
| jEdit | text editor | 5.1.0 | 333/536 | 112.58 | 1517 | 1562 |
| JFreeChart | chart library | 1.2.0 | 210/653 | 100.05 | 401 | 405 |
| JHotDraw | GUI framework | 7.0.6 | 68/284 | 12.01 | 120 | 212 |
| JUnit | unit testing | 4.12 | 69/167 | 6.09 | 138 | 665 |
| Log4j | logging library | 1.2.17 | 62/218 | 10.54 | 118 | 423 |
| Weka | machine learning library | 3.6.11 | 448/1346 | 203.51 | 2121 | 218 |

## 5.2 Multi-view Clustering

In this section, we report a quantitative comparison between single-view and multi-view clusterings on the dataset.

### 5.2.1 Methods and Evaluation Measures

Assessing the quality of the produced clusterings requires constructing an authoritative decomposition. We have adopted the approach used in works [16, 17, 50] by constructing the authoritative decomposition from the package structure of the system. In contrast to the aforementioned approaches where the package structure is flattened, we maintain the tree structure of the package structure, as adopted by [44]. The problem with flattening the package structure is that the hierarchy is lost and hence there is no distinction between modules based on their location in the hierarchy. In other words, the same penalty cost is given to a module that is grouped together with its parent package as a module from another package.

We have opted for hierarchical clustering to perform the cluster analysis. Since hierarchical clustering algorithms work on distance matrices, we convert the produced kernel matrix of data integration $\mathbf{K}$ into a distance matrix, using the following equation, given in [23]:

$$\mathbf{D}^2 = \mathrm{diag}(\mathbf{K})\mathbf{e}^\top + \mathbf{e}(\mathrm{diag}(\mathbf{K}))^\top - 2\mathbf{K} \tag{1}$$

where $\mathrm{diag}(\mathbf{K})$ is a column vector containing the diagonal elements, $\mathbf{e}$ is a column vector of all ones, and the resulting $\mathbf{D}^2$ is the squared distances between all pairs of elements.

For evaluating the authoritativeness of produced hierarchical clusters, we report the path-based difference metric (PD) [34]. For this metric, a lower value indicates better clustering quality. The path-based difference metric is based on the difference between the length of paths of any two leaves between the two trees. This metric has shown to perform well to understand how much the structure of the package structure is preserved, when compared to a baseline tree structure.

### 5.2.2 Experiment Setup

To make a quantitative comparison between single-view and multi-view clustering, we first run all the possible configuration of kernel functions, as outlined in Table 2 in each domain, followed by performing hierarchical clustering on its corresponding distance matrix. The model that yields the best result in each view (considered as the results of single-view clustering) is picked to perform multi-view clustering, as described in Section 3.

The co-training algorithm was run for 50 iterations. Upon completion of the co-training, the spectral embedding of each view were concatenated followed by application of the hierarchical clustering algorithm. If there is prior knowledge on the view informativeness, the most informative view can be used for clustering.

### 5.2.3 Results and Discussion

In Table 4, we make a comparison between different approaches based on the PD score. The grey cell corresponds to the best result for each system. Based on the results of experiments, we make the following observations:

**Best Kernel Functions**    As shown in Table 4, the bag-of-words kernel $K_{BoW}$ outperforms other kernels considered in this paper for semantic similarity. Although the bag of words approach to building the lexical vision of the source code loses the structure in the text, it is a suitable model when dealing with a source code corpus, as in the natural language processing domain. The exponential diffusion kernel $K_{ED}$ is a good choice for inducing a similarity matrix on the module dependency graph, giving the best result in 9 out of 10 cases. We have opted for a polynomial kernel $K_{Poly}$ of order 1 (same as linear kernel) to build the similarity matrix of the software units based on their change history.

**Comparison of Single-view and Multi-view Clustering**    First, for single-view clustering, the structural dependency and the semantic similarity are relatively reliable sources of information, compared to the evolutionary coupling. The evolutionary coupling in project Apache Hadoop is almost 5 times worse than its lexical coupling, indicating the low quality of the evolutionary information in this project.

Second, multi-view clustering techniques establish a lower bound on the performance of the cluster analysis. In all cases without exception, we have observed that the produced decomposition is at least of higher quality than one of the views considered in this work. Hence, in absence of information about the quality of each view, one can integrate different views to produce a representative modularization of the software system.

Third, in 9 out of 10 cases, through multi-view clustering, we have obtained results that are more authoritative than any of the single-view encodings. This observation confirms our hypothesis that by combining different sources of information, we can not only guarantee a lower bound on the performance of clustering but improve it by some margin.

Finally, due to incompatibility of views, the co-training algorithm may not converge. In appendix C, we show how the co-training converges for JFreechart after 30 iter-

■ **Table 4** The comparison of single-view and multi-view clusterings

| System | Single-view Clustering | | | | | | Multi-view Clustering | | |
|---|---|---|---|---|---|---|---|---|---|
| | Struct. | Best K | Evol. | Best K | Lexical | Best K | MKL | Co-training | KCCA |
| Ant | 619 | $\mathbf{K}_{LED}$ (15) | 2198 | $\mathbf{K}_{Poly}$ (1) | 671 | $\mathbf{K}_{BoW}$ | 619 | 618 | 629 |
| Hadoop | 976 | $\mathbf{K}_{ED}$ (10) | 5156 | $\mathbf{K}_{Poly}$ (1) | 1008 | $\mathbf{K}_{BoW}$ | 1072 | 938 | 999 |
| JDT Core | 758 | $\mathbf{K}_{ED}$ ($10^{-1}$) | 2069 | $\mathbf{K}_{Poly}$ (1) | 830 | $\mathbf{K}_{Spec}$ (5) | 805 | 733 | 749 |
| JDOM | 262 | $\mathbf{K}_{ED}$ ($10^{2}$) | 739 | $\mathbf{K}_{Poly}$ (1) | 276 | $\mathbf{K}_{BoW}$ | 319 | 261 | 263 |
| jEdit | 2803 | $\mathbf{K}_{ED}$ (1) | 19028 | $\mathbf{K}_{Poly}$ (1) | 3340 | $\mathbf{K}_{Spec}$ (2) | 3102 | 2565 | 2519 |
| JFreeChart | 1605 | $\mathbf{K}_{ED}$ ($10^{-2}$) | 3832 | $\mathbf{K}_{Poly}$ (1) | 1635 | $\mathbf{K}_{BoW}$ | 2042 | 1469 | 1545 |
| JHotDraw | 264 | $\mathbf{K}_{LED}$ (10) | 516 | $\mathbf{K}_{Poly}$ (1) | 260 | $\mathbf{K}_{BoW}$ | 266 | 242 | 246.0 |
| JUnit | 249 | $\mathbf{K}_{ED}$ ($10^{-5}$) | 612 | $\mathbf{K}_{Poly}$ (1) | 274 | $\mathbf{K}_{Spec}$ (5) | 253 | 244 | 247 |
| Log4j | 189 | $\mathbf{K}_{ED}$ ($10^{-5}$) | 341 | $\mathbf{K}_{RBF}$ (10) | 196 | $\mathbf{K}_{BoW}$ | 188 | 191 | 217 |
| Weka | 3779 | $\mathbf{K}_{ED}$ ($10^{-2}$) | 46113 | $\mathbf{K}_{Poly}$ (1) | 4168 | $\mathbf{K}_{BoW}$ | 5791 | 4104 | 3846 |

ations, whereas in the case of JHotDraw, the convergence does not take place. In general, kernel addition performs very well on compatible views, however, in case of incompatible views, co-training outperforms multiple kernel learning techniques by forcing the views to be as consistent as possible.

### 5.3 Collaborative-Recommendation

In this section, we conduct a quantitative comparison of collaborative recommendation in single-view and multi-view settings.

#### 5.3.1 Methods and Evaluation Measures

We use the "indirect method", as given in [23], to test the kernels for single-view and multi-view recommendation. This method is based on a nearest-neighbour technique that requires a measure of closeness or similarity. The nearest-neighbour scoring algorithm is one of the simplest methods for general classification tasks [19]. The k-nearest-neighbour technique generalizes the idea by taking into account the k nearest data points to decide on class membership. The general idea is that the pattern of items information for an object can be discovered by examining the patterns in the k nearest examples.

The indirect method for item information suggestion is performed by applying a two step process:

1. compute similarity between the object for which recommendations are to be given, and the rest of the objects.

2. use the item information in the k-nearest objects to compute recommendations.

The presence of an item information $i_0$ for a software unit $m_0$ is computed as follows:

$$\text{pred}(m_0, i_0) = \frac{\Sigma_{j=1}^{k} \text{sim}(m_0, m_j) w_{m_j, i_0}}{\Sigma_{j=1}^{k} \text{sim}(m_0, m_j)} \qquad (2)$$

where $w_{m_j, i_0}$ is 1, if information item $i_0$ is a member of $m_j$, and 0 otherwise. The higher the value of $\text{pred}(m_0, i_0)$, the stronger the possibility that $i_0$ exists in $m_0$.

When applying the indirect method, we need to find the top k software units, from the ranked list of modules that are closest to the target module. For each kernel similarity scoring algorithm, we systematically vary k in the range 10 to 100. Parameter k is tuned by using an internal cross-validation.

Link prediction evaluation involves comparing a binary label (whether or not there is an edge) with a real-valued predicted score. There are a variety of methods for evaluation, including fixed-threshold methods such as F1-score, and variable-threshold methods such as the area under the Receiver Operating Characteristic (ROC) curve (AUC), and the area under the Precision-Recall (PR) curve (PRAUC). Due to extreme class imbalance in link prediction task, as only a fraction of pairs of node form edges, it is recommend to use PR curves and its AUC for evaluating link prediction rather than ROC AUC score [53].

To evaluate the performance of the kernel algorithms as well as to compare single-view against multi-view algorithms, we apply the PRAUC and maximum F1-score. The F1-score is the harmonic mean of precision and recall, and the maximum F1 score is the maximum value obtainable for different cut-points.

### 5.3.2 Experiment Setup

The performance of the various single-view and multi-view algorithms is evaluated by applying a standard nested (or double) cross-validation. The internal cross validation involves selecting the kernel and its parameters that gives the best result in single-view and multi-view settings, as well as the choice of the number of nearest neighbours, i.e. k. The evaluation is performed with an internal nine-fold cross validation and performance is averaged on an external ten-fold cross-validation.

Since we are conducting a binary classification, i.e. predicting whether a feature item for an object of interest in a view exists or not, we need to provide binary feature matrices as our input data. We treat this problem as the link prediction in the graph, where the task is to predict missing object-features in the binary matrix. The change history and call graph are binary in nature. In call graph, modules correspond to nodes of the graph and links correspond to existence of 'call' from one module to the other.

In change history, modules and transactions correspond to nodes of the (bipartite) graph, and 'isCommittedIn' links appear as edges connecting the corresponding nodes. For lexical membership, we first perform topic modeling to reduce the set of terms into topics (concepts), so recommendations can be given in terms of high-level concepts.

There are various techniques for topic modeling including Latent Dirichlet Allocation (LDA) [13], Latent Semantic Indexing (LSI) [20] and Nonnegative Matrix Factorization (NMF) [27]. NMF is an unsupervised method that can perform dimensionality reduction as well as clustering simultaneously. NMF reduces the document term matrix in terms of two nonnegative matrices, $D \approx WH$, one corresponding to document-topic relations $W$, and the other topic-term coefficients $H$. Various topic models have been applied to software to extract concepts from the source code [31, 43]. We have opted for NMF as the document-topic relations are naturally very sparse, and correspond to links between documents and topics. To convert module-topic into a binary matrix, the weight of an edge is set to 1 if there is a link between the corresponding items and to 0 otherwise.

### 5.3.3 Results and Discussion

In Tables 5, 6, and 7, we make a comparison between single-view and multi-view approaches for the call graph, change history and topic membership, respectively. The grey cell corresponds to the best result for each system.

**Call Graph Recommendation**   As demonstrated, $\mathbf{K}_{ED}$ emits the best result for the single-view recommendation. In almost all cases, the multi-view setting gives the best result. MKL algorithm outperforms other multi-view learning algorithms by a large margin. On average, the performance is improved by about 50%.

**Change History Recommendation**   Both the Gaussian and polynomial kernels with a small value for their parameters give the best result for this domain. In most cases, MKL outperforms single-view recommendations, but co-training and KCCA give results that are poor compared to single-view setting.

**Topic Membership Recommendation**   Since the topic membership is a binary matrix, both linear kernel and bag-of-words kernel are equivalent and as shown, the linear and small polynomial kernels perform the best. In general, the multi-view configuration outperforms the single-view setting with kernel addition giving the best result.

### 5.3.4 Refactoring Recommendation

Most of the refactoring operations are simple source code transformations aimed at increasing source code comprehension, while preserving the behavior of the system. The goal of refactorings are to correct overall architecture of the software to reflect the developers' perspective of semantic coherency in each view (as demonstrated in the well-modularized software system). Refactorings can be performed at different abstraction levels. Many development environments come with common fine-granular refactoring operations such as 'Rename', 'Move Method', and 'Extract Method'. Recommendations given in this paper are at the artefact level (i.e. classes and files). Here,

■ **Table 5** The comparison of single-view and multi-view CF-based recommendation for call graph

| System | Single-view | | | Multi-view | | | | | |
|---|---|---|---|---|---|---|---|---|---|
| | Structural | | Best K | MKL | | Co-training | | KCCA | |
| | PRAUC | maxF1 | | PRAUC | maxF1 | PRAUC | maxF1 | PRAUC | maxF1 |
| Ant | 0.03 | 0.1 | $K_{ED}$ (1) | 0.08 (+142%) | 0.18 (+68%) | 0.04 (+12%) | 0.12 (+12%) | 0.04 (+5%) | 0.12 (+11%) |
| Hadoop | 0.03 | 0.13 | $K_{LED}$ (1) | 0.06 (+86%) | 0.17 (+29%) | 0.03 (−17%) | 0.1 (−27%) | 0.02 (−35%) | 0.09 (−33%) |
| JDT Core | 0.04 | 0.14 | $K_{ED}$ (100) | 0.1 (+148%) | 0.26 (+83%) | 0.07 (+86%) | 0.18 (+27%) | 0.08 (+97%) | 0.2 (+44%) |
| JDOM | 0.05 | 0.16 | $K_{ED}$ (1) | 0.08 (+49%) | 0.19 (+21%) | 0.06 (+10%) | 0.17 (+10%) | 0.06 (+18%) | 0.14 (−9%) |
| jEdit | 0.07 | 0.15 | $K_{ED}$ (1) | 0.08 (+16%) | 0.17 (+10%) | 0.05 (−20%) | 0.13 (−12%) | 0.03 (−62%) | 0.08 (−48%) |
| JFreeChart | 0.01 | 0.05 | $K_{LED}$ (1) | 0.11 (+854%) | 0.22 (+321%) | 0.01 (−36%) | 0.04 (−26%) | 0.02 (+59%) | 0.08 (+59%) |
| JHotDraw | 0.02 | 0.07 | $K_{ED}$ (0.1) | 0.03 (+49%) | 0.1 (+50%) | 0.03 (+44%) | 0.1 (+46%) | 0.05 (+123%) | 0.15 (+112%) |
| JUnit | 0.02 | 0.08 | $K_{ED}$ (1) | 0.02 (−1%) | 0.08 (−2%) | 0.04 (+80%) | 0.11 (+26%) | 0.03 (+43%) | 0.1 (+17%) |
| Log4j | 0.03 | 0.1 | $K_{ED}$ (0.1) | 0.1 (+200%) | 0.2 (+92%) | 0.04 (+14%) | 0.12 (+20%) | 0.04 (+6%) | 0.13 (+26%) |
| Weka | 0.11 | 0.21 | $K_{ED}$ (1) | 0.07 (−40%) | 0.17 (−17%) | 0.07 (−41%) | 0.17 (−19%) | 0.07 (−39%) | 0.17 (−17%) |

we are more interested in what needs to change to make the system more semantically coherent, as compared to recommending how the actual refactoring should be performed. In the following, we outline how high-level recommendations proposed in this paper can translate into coarse-grained recommendations in the context of the jEdit project. This recommender system can be integrated in a development environment and assist developers by giving recommendations about the module that is currently being viewed and edited.

In the rest of this section, we assume the absence of a link cannot be taken for granted (i.e. reliability of absence of a link), and try to predict new links between the nodes. We conduct the evaluation as outlined in the previous section to produce a prediction score, and use 0.5 as a threshold to predict existence of a link. A cross-validation can be performed to find the optimal choice of cut threshold in each view.

**Structural relationships** In an object-oriented software system, classes are coupled with each other by different types of relationships (e.g., method calls, inheritance, etc.). In this paper, we have restricted ourselves to method invocations, and recommendations involve predicting existence or absence of a call dependency between

**Table 6** The comparison of single-view and multi-view CF-based recommendation for change history

| System | Single-view | | Best K | Multi-view | | | | | |
|---|---|---|---|---|---|---|---|---|---|
| | Evolutionary | | | MKL | | Co-training | | KCCA | |
| | PRAUC | maxF1 | | PRAUC | maxF1 | PRAUC | maxF1 | PRAUC | maxF1 |
| Ant | 0.1 | 0.22 | $K_{Poly}$ (2) | 0.21 (+118%) | 0.26 (+16%) | 0.02 (−81%) | 0.09 (−60%) | 0.08 (−20%) | 0.18 (−19%) |
| Hadoop | 0.05 | 0.16 | $K_{Poly}$ (2) | 0.15 (+179%) | 0.22 (+34%) | 0.04 (−22%) | 0.13 (−22%) | 0.09 (+61%) | 0.18 (+12%) |
| JDT Core | 0.18 | 0.27 | $K_{RBF}$ ($10^{-5}$) | 0.21 (+14%) | 0.31 (+13%) | 0.12 (−32%) | 0.22 (−17%) | 0.16 (−9%) | 0.25 (+6%) |
| JDOM | 0.2 | 0.37 | $K_{Poly}$ (5) | 0.38 (+86%) | 0.43 (+18%) | 0.05 (−73%) | 0.15 (−60%) | 0.2 (−1%) | 0.35 (−4%) |
| jEdit | 0.15 | 0.23 | $K_{Poly}$ (2) | 0.19 (+33%) | 0.24 (+5%) | 0.08 (−42%) | 0.19 (−16%) | 0.1 (−31%) | 0.2 (−11%) |
| JFreeChart | 0.38 | 0.45 | $K_{Poly}$ (4) | 0.4 (+4%) | 0.44 (−2.8%) | 0.09 (−75%) | 0.2 (−57%) | 0.13 (−66%) | 0.22 (−51%) |
| JHotDraw | 0.48 | 0.54 | $K_{RBF}$ (1) | 0.32 (−34%) | 0.39 (−28%) | 0.2 (−57%) | 0.33 (−40%) | 0.3 (−37%) | 0.39 (−28%) |
| JUnit | 0.58 | 0.64 | $K_{Poly}$ (5) | 0.61 (+5%) | 0.63 (−2%) | 0.03 (−95%) | 0.07 (−89%) | 0.15 (−75%) | 0.24 (−63%) |
| Log4j | 0.22 | 0.3 | $K_{Poly}$ (2) | 0.24 (+12%) | 0.3 (+1%) | 0.08 (−62%) | 0.19 (−36%) | 0.13 (−38%) | 0.23 (−23%) |
| Weka | 0.32 | 0.4 | $K_{RBF}$ (1) | 0.27 (−14%) | 0.38 (−5%) | 0.22 (−30%) | 0.37 (−8%) | 0.26 (−16%) | 0.37 (−8%) |

two classes. The following is an example of recommendation for call graph given for 'XMLUtilities' class:

> *org.gjt.sp.util.XMLUtilities* should make a call to *org.gjt.sp.util.IOUtilities*.

After examining these two classes, the recommendation is quite sensible. XMLUtilities is a wrapper of functions for parsing XML files with other utilities involving I/O. These I/O functionalities can be delegated to IOUtilities, resulting in separation of concerns between these two utility classes.

**Change history** The change history allows us to give suggestions about potentially relevant source code to a developer performing a modification task. The recommendations predict and suggest likely changes as part of a single commit, and prevent errors due to incomplete changes. There have been several works [54, 55] that employ pattern mining techniques such as association rule mining to mine change history to help a developer identify relevant source code units under modification.

Assuming a new commit is to be made with a file staged, here 'PluginJAR.java', we try to make a prediction as to what other modules need to be co-comitted.

■ **Table 7** The comparison of single-view and multi-view CF-based recommendation for topic membership

| System | Single-view | | | Multi-view | | | | | |
|---|---|---|---|---|---|---|---|---|---|
| | Topic | | Best K | MKL | | Co-training | | KCCA | |
| | PRAUC | maxF1 | | PRAUC | maxF1 | PRAUC | maxF1 | PRAUC | maxF1 |
| Ant | 0.44 | 0.47 | $\mathbf{K}_{Poly}$ (3) | 0.43 (−1%) | 0.5 (+7%) | 0.36 (−18%) | 0.45 (−4%) | 0.37 (−15%) | 0.48 (+3%) |
| Hadoop | 0.19 | 0.3 | $\mathbf{K}_{Poly}$ (2) | 0.29 (+50%) | 0.39 (+31%) | 0.23 (+21%) | 0.33 (+8%) | 0.17 (−14%) | 0.26 (−12%) |
| JDT Core | 0.19 | 0.32 | $\mathbf{K}_{Poly}$ (3) | 0.24 (+24%) | 0.4 (+26%) | 0.21 (+11%) | 0.35 (+8%) | 0.22 (+20%) | 0.37 (+16%) |
| JDOM | 0.3 | 0.36 | $\mathbf{K}_{Poly}$ (3) | 0.32 (+7%) | 0.39 (+8%) | 0.29 (−4%) | 0.41 (+14%) | 0.33 (+11%) | 0.45 (+23%) |
| jEdit | 0.21 | 0.3 | $\mathbf{K}_{Poly}$ (4) | 0.34 (+63%) | 0.39 (+30%) | 0.35 (+68%) | 0.37 (+24%) | 0.24 (+17%) | 0.33 (+10%) |
| JFreeChart | 0.13 | 0.22 | $\mathbf{K}_{Poly}$ (3) | 0.29 (+121%) | 0.38 (+71%) | 0.18 (+38%) | 0.26 (+15%) | 0.14 (+4%) | 0.21 (−5%) |
| JHotDraw | 0.17 | 0.29 | $\mathbf{K}_{Poly}$ (2) | 0.21 (+24%) | 0.3 (+3%) | 0.19 (+11%) | 0.27 (−6%) | 0.16 (−5%) | 0.28 (−6%) |
| JUnit | 0.2 | 0.34 | $\mathbf{K}_{Poly}$ (2) | 0.17 (−14%) | 0.32 (−5%) | 0.23 (+13%) | 0.35 (+5%) | 0.22 (+8%) | 0.35 (+2%) |
| Log4j | 0.2 | 0.35 | $\mathbf{K}_{Poly}$ (3) | 0.25 (+24%) | 0.41 (+17%) | 0.3 (+48%) | 0.4 (+14%) | 0.27 (+34%) | 0.42 (+20%) |
| Weka | 0.29 | 0.36 | $\mathbf{K}_{Poly}$ (4) | 0.34 (+18%) | 0.39 (+8%) | 0.32 (+10%) | 0.39 (+8%) | 0.31 (+7%) | 0.37 (+2%) |

> *org/gjt/sp/jedit/PluginJAR.java* should be committed with *org/gjt/sp/jedit/Edit-Plugin.java* and *org/gjt/sp/jedit/msg/PluginUpdate.java*.

**Lexical information**    Using the topic membership recommendation, it is possible to give suggestions to refine both what functionalities should the source code unit provide as well the identifier naming, to make it more aligned with the semantic role the unit plays. As proposed by Abebe and Tonella [1], recommendation techniques can exploit ontological concepts and relations between identifiers to support these refactorings.

We have performed NMF to extract 7 topics, over a total of 2556 words. Table 8 depicts the top 10 keywords for each topic and its associated concept, which we have manually inferred by examining those words.

> *org.gjt.sp.jedit.textarea.TextArea* should cover more of topic 4 (i.e. *buffer, view, path, file, directori*).

■ **Table 8** The 7 topics extracted from jEdit bag-of-words using NMF

|  | Topic 1 | Topic 2 | Topic 3 | Topic 4 | Topic 5 | Topic 6 | Topic 7 |
|---|---|---|---|---|---|---|---|
| **Concept** | **Common** | **Plugins** | **BeanShell** | **FileSystem** | **TextArea** | **Expression** | **GUI** |
| | cur | plugin | jjtree | buffer | line | rule | line |
| | move | jar | token | view | caret | context | font |
| | state | cach | scope | path | offset | tag | col |
| | pos | entri | scan | file | select | block | grid |
| **Top-10** | stop | tabl | node | directori | scroll | match | layout |
| **keywords** | check | resourc | consum | properti | screen | opcod | color |
| | start | path | express | config | word | instruct | height |
| | input | uri | pars | vfs | posit | parser | width |
| | string | model | liter | filter | visibl | keyword | rowspan |
| | read | file | lparen | browser | text | insn | colspan |

The topic 4 corresponds to file system management functionalities in jEdit. The prediction entails recommending to the developer to perform operations or concepts that correspond to the topic.

### 5.4 Cross-modal Retrieval

In this section, we explore cross-modal retrieval using shared subspace learning to evaluate the task of source code search in the context of jEdit project. We employ kernel CCA to find a set of projections from various views into a shared semantic space. The hypothesis is that finding such correlations between views will reveal the underlying semantics. The directions carry information about the latent concepts that was the factor behind generation of various views, i.e. the textual content of the source code, module dependencies and how modules were co-committed. To perform a cross-modal search task, we first project the query $\mathbf{q}$ from one domain into the shared subspace, followed by finding the nearest neighbors in the subspace. This can serve as part of a natural language search engine for source code search as well as code synthesis to construct boilerplates or templates based on similar modules (as disccused in CF-based recommender systems).

For this task, the query we are interested in is: '**The class that handles search dialog**'. The kernel used in this evaluation to compute the similarity between the query and software modules in the textual view is $\mathbf{K}_{BoW}$. To find a representative vectorization of the query text, we decompose the query into a vector of features comprising of stemmed words, weighed by tf-idf weighing score. The resulting query feature vector comprises of the following words: $[handl, search, dialog]$. We proceed by computing the similarity between query feature vector and software modules. The top-10 nearest neighbors (i.e. most similar) are returned as the result of the search query. For cross-modal setting, we perform kernel CCA on the jEdit project, projecting the data across views into a shared 2-dimensional subspace. The top-10

■ **Figure 2**   The shared subspace of query and the softwatre entities

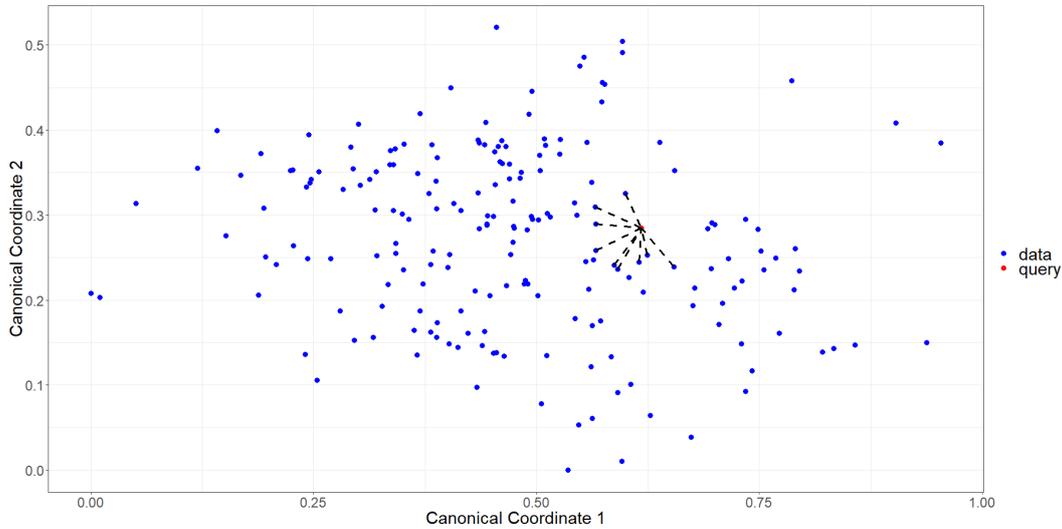

nearest neighbors are returned as the result of the search query. For computing the similarity in other views, we have used $\mathbf{K}_{ED}$ with $\alpha = 1$ for structural dependency graph and the linear kernel $\mathbf{K}_{Poly}$ with $d = 1$ for change history. For more information on how to perform cross-modal retrieval using kernel CCA, please refer to appendix D. Figure 2 depicts the query and modules in the shared subspace. After examination of the source code, we found 4 classes that handle different stages of lifecycle of search dialog:

VFSBrowser, SearchBar, SearchAndReplace, HyperSearchResults

Table 9 gives the top-10 ranked results from uni-modal and cross-modal searches for the aforementioned query. Classes with 'Search' in their names (and in their content) are ranked high in both search methods, however 'VFSBrowser' is only present in the cross-modal setting. Both methods miss out on 'HyperSearchResults' class. This demonstrates the usefulness of cross-modal retrieval for code search.

### 5.5  Threats to Validity

The generalizability of our findings in this paper is limited by the restricted set of projects comprising of 10 open source Java projects. We believe our techniques should be evaluated with larger systems from different programming paradigms to see if improvement still occurs. Another major threat to validity of this research is constructing the ground truth for evaluating the clustering results. We have relied on the package structure to build the authoritative decomposition. Although there is no solid agreement on what constitutes a good architecture [45], we believe the least our approach can accomplish is to retrieve/construct the package structure of a software system. On the other hand, we have opted for projects which are well-engineered or have gone through a migration, and hence, the package structure is a good indicator

■ **Table 9** Top-10 ranked results from uni-modal and cross-modal retrieval

|  | **Uni-modal Search** | **Cross-modal Search** |
|---|---|---|
| | SearchDialog | SearchAndReplace |
| | SearchBar | SearchBar |
| | SearchSettingsChanged | SearchDialog |
| | SearchAndReplace | SearchSettingsChanged |
| **Top**-10 | ServiceListHandler | GUIUtilities |
| **classes** | ActionListHandler | DynamicContextMenuService |
| | JJTParserState | VFSFileFilter |
| | BshClassManager | VFSBrowser |
| | ErrorListDialog | VFSFileChooserDialog |
| | RolloverButton | VFSManager |

of the architecture of the system. Furthermore, we eliminate packages with fewer than 4 classes and manually split packages with more than 40 classes in them. Eliminating small and large packages ensure that the oracle decomposition itself doesn't exhibit extreme distribution, and is uniformly grouped. Finally, preserving the hierarchical structure of the package structure ensures topological information is not lost, and hence a better ground truth to evaluate the cluster analysis.

## 6 Conclusion and Future Work

We have adopted multi-view learning for modularization and recommendation for software systems as well as source code search, where different sources of knowledge contribute to what a 'good' modularity of a software system should be, what item information can be 'best' recommended for a software unit, and what the semantic meaning of a query is, respectively. The observation we have made through conducting the experiments on 10 relatively large Java open source projects is that incorporating different sources of information relevant to a software system through multi-view learning improves the general quality of software modularization and recommendation. We would like to investigate whether the techniques introduced here could help with modularization and refactoring recommendations for legacy software systems, which brings its own set of complications such as obsolete code and arbitrary naming.

To the best of our knowledge, this is the first time that multi-view learning is employed to address a problem in the software engineering domain. There are many other areas including mining repositories and source code analysis that multi-view learning can be employed to improve results. Based on the techniques introduced in this paper, we want to incorporate other sources of information relevant to software systems such as file ownership and authorship. Also, further study of kernel functions for learning from software artefacts is essential including the choice of kernel functions as well as tuning the parameters. Also, we would like to further investigate the reasons behind the empirical values found for both the kernel functions and their parameters.

## A  Multi-view Learning Approaches

### A.1  Co-training

Co-training is one of the earliest schemes of multi-view learning introduced by Blum and Mitchell [14] for semi-supervised classification. It trains alternatively between the views to maximize the mutual agreement between the distinct views of the unlabeled data. Bickel and Scheffer [12] built on this approach to develop a co-training-based multi-view algorithm to learn the model parameters by alternating between the views and subsequently estimate the cluster assignments. Figure 3 sketches the co-training-based algorithms for learning on multi-view data. The idea behind the co-training approach to multi-view learning is to train each view with the hypothesis of the other view until views converge to a stationary point. In case where the views are incompatible, there is no guarantee of convergence [30].

In Kumar and Daume's work [30], a co-training approach is used to perform clustering within one view and transfer the constraints on the similarity graph of the other view. This process is repeated for some iterations. We have implemented this co-training approach to perform multi-view learning.

### A.2  Multiple Kernel Learning

Multiple kernel learning (MKL) involves incorporting multiple kernels corresponding to different representation of data into a single kernel matrix. Kernel functions on different modes of data denote different notion of similarity in each view, which

■ **Figure 3** The scheme for co-training in multi-view setting can be in practice contrasting. MKL achieves multi-view learning by fusing different sources of information into one by means of linear or non-linear combination. Figure 4 depicts the multiple kernel learning in multi-view data.

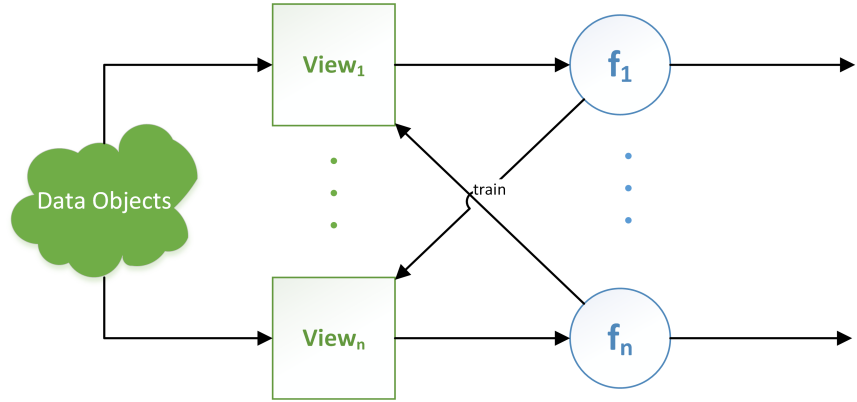

■ **Figure 4** The scheme for multiple kernel learning in multi-view setting

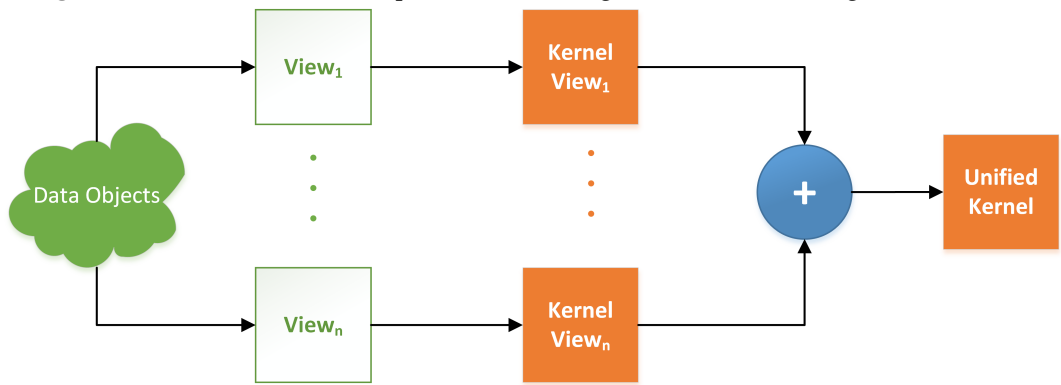

For simplicity, we only consider one type of MKL for performing multi-view learning:

**Kernel Addition** In general, kernel addition reduces to concatenation of features in the reproducing kernel Hilbert space (RKHS). In spite of its simplicity, this method is shown to outperform many other sophisticated techniques.

### A.3 Subspace Learning

Subspace learning for multi-view learning tries to obtain a latent representaiton shared between different views. The assumption is that all the views were generated from a single latent subspace. The jointly learned subspace can then be used to perform learning tasks such as classification and clustering. As the subspace has a lower dimension than any of the views, the problem of 'curse of dimensionality' is resolved. Figure 5 depicts how subspace learning can be used for multi-view learning.

■ **Figure 5** The scheme for shared subspace learning

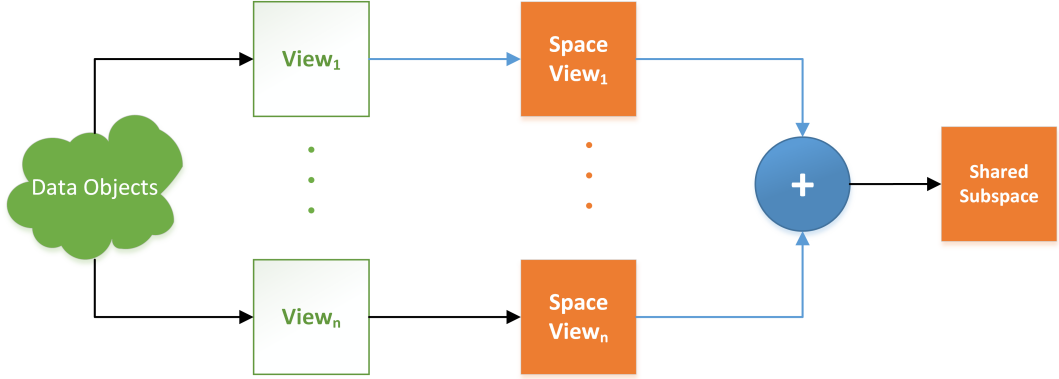

The widely-used dimensional reduction technique for single-view learning is Principle Component Analysis (PCA). The multi-view variant of this technique is called Canonical Correlation Analysis (CCA) [26]. CCA tries to explore the basis vectors of two set of variables, $\mathbf{V}_1$ and $\mathbf{V}_2$, by mutually maximizing the correlations between the projections onto the set of basis vectors. The main assumption is that the representations in these two spaces share common information that is reflected in correlations between them. CCA learns two low-dimensional subspaces $\mathbf{S}_1 \subset \mathbf{V}_1$ and $\mathbf{S}_2 \subset \mathbf{V}_2$ of the original spaces that maximizes the correlation between them, defined as follows:

$$\max_{\mathbf{W}_1 \neq 0, \mathbf{W}_2 \neq 0} \frac{\mathbf{W}_1^\top \Sigma_{1,2} \mathbf{W}_2}{\sqrt{\mathbf{W}_1^\top \Sigma_{1,1} \mathbf{W}_1} \sqrt{\mathbf{W}_2^\top \Sigma_{2,2} \mathbf{W}_2}} \tag{3}$$

where $\Sigma_{1,1}$ and $\Sigma_{2,2}$ denote the covariance matrices for each of the two variable sets, while $\Sigma_{1,2} = \Sigma_{2,1}^\top$ represent the cross-covariance matrices between them. The above optimization can be solved as a generalized eigenvalue problem:

$$\begin{pmatrix} 0 & \Sigma_{1,2} \\ \Sigma_{2,1} & 0 \end{pmatrix} \begin{pmatrix} \mathbf{W}_1 \\ \mathbf{W}_2 \end{pmatrix} = \lambda \begin{pmatrix} \Sigma_{1,1} & 0 \\ 0 & \Sigma_{2,2} \end{pmatrix} \begin{pmatrix} \mathbf{W}_1 \\ \mathbf{W}_2 \end{pmatrix} \tag{4}$$

The generalized eigenvectors, i.e. $\mathbf{W}_1$ and $\mathbf{W}_2$, determine a set of uncorrelated canonical components, whereas the corresponding eigenvalues $\lambda$ denote the explained correlation. The first $d$ canonical components $\{\mathbf{W}_{1,i}\}_{i=1}^d$ and $\{\mathbf{W}_{2,i}\}_{i=1}^d$ define a set of basis vectors that can be used to project $\mathbf{V}_1$ and $\mathbf{V}_2$ into subspaces $\mathbf{S}_1$ and $\mathbf{S}_2$, respectively. Since these two set of vectors are coordinates in two isometric subspaces, they can be combined to obtain a shared common subspace $\mathbf{S}$ by overlaying $\mathbf{S}_1$ and $\mathbf{S}_2$.

The linear methods such as CCA fail to fully capture the properties of the data which exhibit non-linearities. To deal with the non-linearities, kernel method has been successfully employed in many real-world applications (e.g. Kernel Principal Component Analysis). The kernel variant of CCA is proposed that overcomes this problem by first projecting each dimension into higher dimensional space using a kernel function, followed by applying linear CCA in that space.

## B    Kernel functions

In the following, we consider three classes of kernels: 1) kernels that operate on vector spaces, 2) kernels over strings, and 3) graph kernels.

### B.1  Kernels for Vector Spaces

We outline some of kernels that are used for vectors space models.

### B.1.1  Polynomial Kernel

The most basic example of a kernel for vector spaces with inner product is the *linear kernel* defined as $\mathbf{K}_L(x, y) = \langle x, y \rangle$. The generic form of the linear kernel is the polynomial kernel defined as:

$$\mathbf{K}_{Poly}(\mathbf{x}_i, \mathbf{x}_j) = (\mathbf{x}_i^\top \mathbf{x}_j + r)^d \tag{5}$$

where $d$ is a positive integer denoting the degree of the polynomial and $r$ is a non-negative real number, trading off the influence of higher-order versus lower-order terms in the polynomial. We have chosen the scale $r = 0$ throughout our experiments.

### B.1.2  RBF Gaussian Kernel

Another important kernel is the RBF Gaussian kernel which defines feature space in terms of infinite number of dimensions, defined as:

$$\mathbf{K}_{RBF}(\mathbf{x}_i, \mathbf{x}_j) = \exp(-\frac{\|\mathbf{x}_i - \mathbf{x}_j\|^2}{\sigma^2}) \tag{6}$$

where $\sigma > 0$ is a free parameter. A Gaussian kernel with very high-dimension, as is the case with say the document-term matrix, is known to produce unreliable results. Hence, we first project the vector space into a lower dimensional space using Latent Semantic Analysis (LSA) [20] and then compute the similarities in this space. In general, the Gaussian kernel is a reasonable choice when dealing with non-linear data, however when the feature space is too large, one may opt for the linear kernel.

### B.1.3  Bag-of-words Kernel

The *vector space model* (VSM) is a well-known model used to express documents for information retrieval. Given a set of words, and a set of documents, the bag-of-words model treats each document as a unordered collection of words. The weights of terms are usually normalized by the well-known term-frequency-inverse-document-frequency weighing mechanism (tf-idf) to punish common terms while giving more weight to rare words in the corpus. The bag-of-words kernel is then used to measure the similarity between documents. The bag-of-words kernel is analogous to the cosine similarity measure which computes the cosine of the angle $\theta$ between two documents $\mathbf{x}_i$ and $\mathbf{x}_j$ in the VSM.

$$\mathbf{K}_{BoW}(\mathbf{x}_i, \mathbf{x}_j) = \frac{\mathbf{x}_i^\top \mathbf{x}_j}{\|\mathbf{x}_i\|\|\mathbf{x}_j\|} \tag{7}$$

## B.2 String Kernels

Although the bag-of-words kernel works well in natural language processing, it is an open question if the same techniques employed for feature extraction and normalization perform in the context of source code analysis. String kernels [32] on the other hand, are based on matching of substrings in the text of documents. An advantage of string kernels is that it can be applied directly to the text of the document without the need for extraction and normalization of features of documents. The string kernels are efficiently computed using a suffix tree[5].

Consider an alphabet $\Sigma$ and its Kleene closure $\Sigma^*$ that represents the set of non-empty strings defined over the alphabet $\Sigma$ augmented with the empty string $\epsilon$. The string kernel is then defined as:

$$\mathbf{K}_S(\mathbf{x}_i, \mathbf{x}_j) = \sum_{s \in \mathbf{x}_i, s' \in \mathbf{x}_j} \lambda_s \delta_{s,s'} = \sum \text{num}_s(\mathbf{x}_i) \, \text{num}_s(\mathbf{x}_j) \lambda_s \qquad (8)$$

where $\text{num}_s(\mathbf{x}_i)$ denotes the number of occurrences of string $s$ in $\mathbf{x}_i$ and $\lambda_s$ controls the weight decay of string $s$. A string kernel can also be controlled by the parameter, "length of a subsequence $k$". Different configurations of these parameters give rise to different types of string kernel. In this paper, we will only examine three different types of string kernels:

- Constant ($\mathbf{K}_{Cons}$): all common substrings are matched with equal weighing.
- p-spectrum ($\mathbf{K}_{Spec}$): only common substrings of length $p$ is counted. For the experiments, we have fixed the weighing, i.e. $\lambda_s = 1$.
- Exponential decay ($\mathbf{K}_{Exp}$): all common substrings are matched but the substring weight decays as the matching substring gets shorter.

## B.3 Kernel for Graphs

Kernels for graphs incorporate information about the structure of the graph in terms of a kernel. The function of the kernel for a graph must be able to express a global similarity between any two nodes in a graph whereas an adjacency matrix describes the local similarity, i.e. whether two nodes are neighbours. The notion of a graph kernel here denotes point-wise similarity between nodes of a graph, and should not be confused with those kernels used to establish similarity between two different graphs. For a full survey and empirical comparison and evaluation of different kernels on graphs, please refer to [23].

Before we define graph kernels used in this paper, we need to define the laplacian of a graph. For a weighted, undirected graph $G$ with symmetric $w_{ij} > 0$ representing the weight of edges between nodes $i$ and $j$, the laplacian matrix is defined as $\mathbf{L} = \mathbf{D} - \mathbf{A}$, where $\mathbf{D} = \text{diag}(\mathbf{x}_i)$ is the (generalized) degree matrix, with diagonal entries $\mathbf{x}_i = \sum_{j=1}^{n} \mathbf{A}_{ij}$. Since the module dependency graph is directed, we will make the

---

[5] A suffix tree is a data structure that represents the suffixes of a string $S$ in terms of a tree-like structure.

■ **Figure 6**   Convergence of co-training algorithm for JFreeChart and JHotDraw

**(a)** PD scores in different views vs. the number of iterations of co-trained hierarchical clustering for JFreeChart

**(b)** PD scores in different views vs. the number of iterations of co-trained hierarchical clustering for JHotDraw adjacency matrix symmetric using the symmetric operator $\text{sym}(\mathbf{A}) = \frac{\mathbf{A} + \mathbf{A}^\top}{2}$. We will make a comparison between two kernels on a graph, namely the exponential diffusion kernel and the laplacian exponential diffusion kernel.

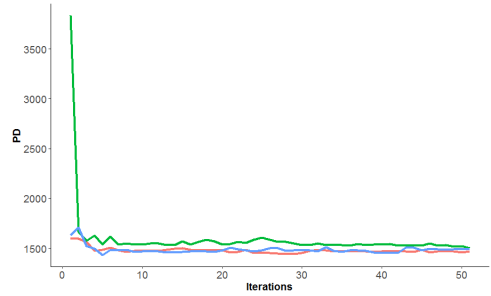
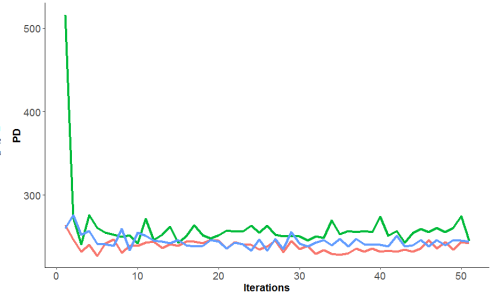

### B.3.1 Exponential Diffusion

Inspired by the physical process of diffusion, several different kernels based on the diffusion model have been proposed. The exponential diffusion kernel is one such example that was originally introduced by Kondor and Lafferty [28]. The diffusion kernel is computed through a power series of the adjacency matrix of the graph. Here, $\alpha$ denotes the rate of "sink" in each node of the graph.

$$\mathbf{K}_{ED} = \sum_{k=0}^{\infty} \frac{\alpha^k \mathbf{A}^k}{k!} = \exp(\alpha \mathbf{A}) \tag{9}$$

### B.3.2 Laplacian Exponential Diffusion

Another diffusion model on graphs, proposed by Smola and [48], is the laplacian exponential diffusion (also known as the heat diffusion kernel) which substitutes the adjacency matrix $\mathbf{A}$ in equation 9 with the laplacian of the graph.

$$\mathbf{K}_{LED} = \exp(-\alpha \mathbf{L}) \tag{10}$$

## C   Convergence of Co-training Algorithm

Figure 6 shows the change in the PD metric of the produced hierarchical clusterings against the number of iterations. For JFreeChart, the algorithm converges after around 30 iterations. On the other hand, for the JHotDraw project, the convergence does not take place, and the value of PD metric keeps varying through the subsequent iterations. As shown, the least informative views do gain alot after the first iteration. In four of the cases, the algorithm does not converge. This observation can be due to the non-compatibility of the views.

### D   Kernel CCA for cross-modal retrieval

Performing cross-modal retrieval or multi-view embedding using kernel CCA is analogous to using a kernel PCA followed by a CCA, which allows us to capture the nonlinearity of the transformation between the components of different views. Hence, finding the nearest neighbors of a query vector $\mathbf{q}$ in the shared canonical space involves projecting the query vector onto the kernel principal axes, followed by mapping it into the shared canonical space. The procedure for computing the nearest neighbors of the query vector using kernel CCA is as follows:

For a kernel matrix $\mathbf{K}$ with $\mathbf{K}_{ij} = k(\mathbf{x}_i, \mathbf{x}_j)$, we first compute the kernel principal components:

1. Center the kernel matrix via the following equation:

$$\mathbf{K}_{\text{centered}} = \mathbf{K} - \mathbf{1}_n\mathbf{K} - \mathbf{K}\mathbf{1}_n + \mathbf{1}_n\mathbf{K}\mathbf{1}_n = (\mathbf{I} - \mathbf{1}_n)\mathbf{K}(\mathbf{I} - \mathbf{1}_n),$$

   where $\mathbf{1}_n$ is a $n \times n$ matrix with all elements equal to $\frac{1}{n}$, and $n$ is the number of data points.

2. Using the eigendecomposition of the centered kernel matrix: $\mathbf{K}_{\text{centered}} = \mathbf{U}\mathbf{S}^2\mathbf{U}^\top$. The kernel principal axes $\mathbf{PC}$ are computed by multiplying each eigenvector by the square root of the respective eigenvalue: $\mathbf{PC} = \mathbf{U}\mathbf{S}$.

For a query vector $\mathbf{q}$ specified in terms of view $v$, we project it onto the kernel principal axes of the corresponding view:

1. For each original data point $\mathbf{x}_i$ and the kernel function $k_v(.,.)$ for view $v$, the query similarity vector is computed by $\mathbf{k}_q = k_v(\mathbf{q}, \mathbf{x}_i)$.

2. Find the projection of the original query in the principal component axes: $\mathbf{q}_{pc} = \mathbf{k}_q\mathbf{U}_v\mathbf{S}_v^{-1}$

The computed query vector $\mathbf{q}_{pc}$ is then projected into the canonical space using its corresponding projection vector $\mathbf{W}_V$, i.e. $\mathbf{q}_{cca} = \mathbf{q}_{pc}\mathbf{W}_v$, as explained in appendix A.3. The nearest neighbors of $\mathbf{q}_{cca}$ are then those data points with smallest distance, for some suitable measure of distance. We have used the euclidean distance to find the nearest neighbors of the query.

## About the authors

**Amir M. Saeidi** is a software engineer at Mendix and a PhD candidate at the Department of Information and Computing Sciences at Utrecht University. He is currently investigating various techniques to facilitate migration of legacy systems in financial domain to SOA. The techniques employed range from static analysis to data analysis to help both with understanding the legacy systems as well as their decomposition. He can be reached at a.m.saeidi@uu.nl. 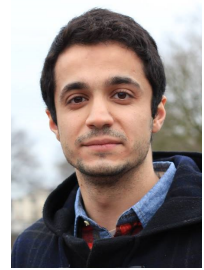

**Jurriaan Hage** is an assistant professor at the Department of Information and Computing Sciences at Utrecht University. His work in programming technology is largely focused on two aspects: the optimisation of functional languages by means of type and effect systems, and type error diagnosis for strongly typed functional languages. He is currently the lead maintainer of the Helium compiler for learning Haskell. Besides these two focus areas, he is also active in programming plagiarism detection, legacy system modernization, and the (soft type) analysis of dynamic languages. He can be reached at j.hage@uu.nl. 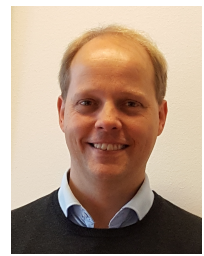

**Ravi Khadka** is currently a technology architecture manager at Accenture. His focus area include legacy software modernization, light-weight architecture, model-driven development (MDD), API management, and micro-service architecture. Khadka received his PhD in computer science from Utrecht University in 2016. His PhD thesis is titled "Revisiting Legacy Software System Modernization". He can be reached at ravi.khadka@gmail.com. 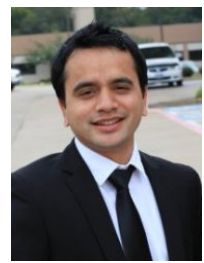

**Slinger Jansen** is an assistant professor at the Department of Information and Computing Sciences at Utrecht University. His research focuses on software product management and software ecosystems, with a strong entrepreneurial component. Jansen received his PhD in computer science from Utrecht University, based on the 2007 work titled "Customer Configuration Updating in a Software Supply Network", PhD thesis Utrecht University. He can be reached at slinger.jansen@uu.nl. 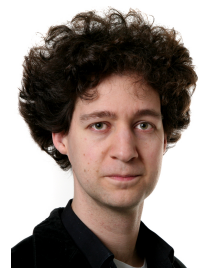